\documentclass[a4paper,11pt]{article}
\pdfoutput=1 

\usepackage{jcappub} 
\usepackage[a4paper, total={7in, 9in}]{geometry}
\usepackage[T1]{fontenc} 
\usepackage{graphicx}
\usepackage{epsf}
\usepackage{bm}
\usepackage{amsmath}
\usepackage{amsfonts}
\usepackage{amssymb}
\usepackage{epstopdf}
\usepackage{natbib}
\usepackage{hyperref}
\usepackage{color}
\usepackage{verbatim}
\usepackage{multirow}
\usepackage{bm}
\usepackage{float}
\usepackage{tabularx}

\definecolor{darkblue}{rgb}{0.0, 0.0, 0.55}
\definecolor{darkred}{rgb}{0.55, 0.0, 0.0}

\usepackage{hyperref}
\hypersetup{
   colorlinks=true, 
    linkcolor=darkblue,
    citecolor=darkblue,
    urlcolor=darkblue}

\newcommand{\Mpl}{M_{\rm Pl}}

\begin{document}

\title{Analytical constraints on gravitational models with a quadratic Weyl tensor\\[0.5em] \footnotesize\texttt{KCL-PH-TH-2025-11 YITP-25-61}}

\author[a]{Benjamin Sutton,}
\author[b]{Antonio de Felice}
\author[a]{and Mairi Sakellariadou}


\affiliation[a]{Theoretical Particle Physics and Cosmology Group, Physics Department,
King’s College London, University of London, \\ Strand, London WC2R 2LS, United Kingdom}
\affiliation[b]{Center for Gravitational Physics and Quantum Information, 
Yukawa Institute for Theoretical Physics, Kyoto University, \\ 606-8502, Kyoto, Japan}

\emailAdd{benjamin.j.sutton@kcl.ac.uk}
\emailAdd{antonio.defelice@yukawa.kyoto-u.ac.jp}
\emailAdd{mairi.sakellariadou@kcl.ac.uk}

\date{\today}

\abstract{We set analytical constraints on the parameter space of models of gravity containing a term quadratic in Weyl curvature $-\alpha C^2$. 
In this class of models, there are four propagating tensorial degrees of freedom, 
two vector degrees of freedom, and two 
scalar degrees of freedom, $\delta_m$ and $\Phi$, 
corresponding to gauge invariant perturbations in the matter density and the gravitational Bardeen potential, respectively. 
We consider the era of matter domination, and requiring that growth of perturbations are recovered in the scalar sector and classical instabilities are eliminated in the vector and tensor sectors, we obtain bounds on the free coupling parameter to the quadratic Weyl curvature term, $10^{14}H^2_0 \lesssim \alpha^{-1} \ll M^2_{\text{cutoff}}$, where $M_{\text{cutoff}}$ is the cutoff scale of the low energy effective field theory, and $\alpha^{-1}$ is proportional to the masses of the additional propagating degrees of freedom.}

\maketitle

\section{Introduction}\label{sec:intro}

The limitations of General Relativity (GR) as an effective field theory (EFT) at high energies have motivated the development of modified gravity (MG) models \cite{Clifton:2011jh}. These models often introduce higher-order curvature terms into the 
gravitational action to resolve current issues in fundamental physics and cosmology, such as the ultraviolet (UV) completion of gravity, the origin of cosmic inflation, and the nature of dark energy and dark matter. 
One such approach, which has gained interest, is the inclusion of a quadratic Weyl tensor term, $-\alpha C^2$, initially proposed by Stelle \cite{Stelle:1977ry}. These models are particularly attractive because they are renormalizable and asymptotically free \cite{Fradkin:1981iu}, providing a potential framework for a more complete theory of gravity. The introduction of higher-order curvature terms in the gravitational Lagrangian generally gives rise to new propagating degrees of freedom (d.o.f.). However, models of modified gravity are often plagued by ghost d.o.f.s, which manifest from derivatives higher than the second order of the metric in the equations of motion (EoM). Despite these potential complications, the linear perturbations of scalar, vector, and tensor fields propagate at the speed of light in a flat Minkowski background \cite{Hindawi:1995an, Bogdanos:2009tn, Hinterbichler:2015soa}, and, in the absence of matter, these ghost d.o.f.s do not induce classical instabilities like Laplacian or tachyonic instabilities \cite{Nelson:2010rt}. This behavior, however, is specific to Minkowski space, and these results are not necessarily valid in the presence of an expanding background or matter sources \cite{DeFelice:2023psw}. Here, we note that the presence of a ghost does not necessarily lead to classical and/or quantum instabilities. For the former instability, the perturbation equations might lead to stable solutions. For the latter, if classical stability holds, then one can avoid vacuum instabilities and keep the theory unitary through the ``fakeon'' prescription \cite{Anselmi:2018kgz}.

In this work, we extend the analysis of quadratic gravity models with a Weyl curvature term in an expanding Friedmann-Lemaître-Robertson-Walker (FLRW) background, during the era of matter domination. In this phase of cosmic evolution, GR predicts that matter density fluctuations grow linearly with the scale factor, and the gravitational potential remains approximately constant over time, which is instrumental in the formation of large-scale structures \cite{Mukhanov:1990me}. Our focus is on the evolution of scalar perturbations during this epoch, with particular emphasis on the Jeans instability.
In an expanding background, when the universe is dominated by a pressureless fluid, these perturbations grow as a power-law \cite{Bardeen:1980kt,Peebles:1980yev}.
Previous work has considered these models in the context of cosmic inflation, specifically for the Starobinsky model \cite{DeFelice:2023psw}, which introduces a quadratic Ricci scalar term. In particular, in that work, it was shown that the Starobinsky inflationary background was made unstable by the presence of the Weyl squared term, at least as long as the mass of the extra modes is below the cutoff scale $M_{\rm cutoff}$ of the theory. It is then important to see whether this same term can spoil the standard phenomenology of $\Lambda$CDM during matter domination and if so, sets constraints on the allowed value of $\alpha$. In other words, by extending the stability considerations to the epoch of matter domination, we explore what regions of the parameter space of these models recover the growth of perturbations and cosmic structures at late times. We find that for $\alpha<0$ (but still $|\alpha|^{-1}\ll M_{\rm cutoff}^2$, so that the extra modes are not too massive), the perturbation modes are classically unstable, and we discard this region of parameter space. For non-negative $\alpha$, stability is achieved, but we can set a lower bound on $\alpha^{-1}$ (proportional as we will see to the masses of the extra degrees of freedom) by requiring that the growth of perturbation does not change significantly ($\Delta P/P\leq 10\%$) the matter power spectrum compared with General Relativity. If $\alpha^{-1}$ is below this lower bound, we find a suppression of the Jeans instability. We obtain $10^{14}H^2_0 \lesssim \alpha^{-1} \ll M^2_{\text{cutoff}}$. This inequality makes sense only if $M_{\rm cutoff}^2\gg 10^{14} H_0^2$, leading to a lower bound on the cutoff scale for the theory. 

The paper is structured as follows. We begin by setting up the model of quadratic gravity in Section \ref{The action of quadratic gravity}. We then consider the dynamics of an expanding FLRW background in quadratic gravity in Section \ref{Dynamics of the Background}. Next, we proceed to discuss the metric perturbations on an expanding FLRW background, and the choice of an appropriate gauge to study the propagation of the modes in Section \ref{Metric perturbations}. Following this, we derive the equations of motion for the propagating d.o.f.s in the vector, tensor, and scalar sectors in Sections \ref{Propagation of the vector modes}, \ref{Propagation of the tensor modes}, and \ref{Propagation of the scalar modes}, respectively. In Section \ref{Conclusions} we discuss the key results and future scope. 

\section{The action of quadratic weyl gravity} \label{The action of quadratic gravity}
The EFT of quadratic gravity 
we consider is \cite{Stelle:1977ry} 
\begin{equation}
    S = \frac{\Mpl^2}2\int d^4 x\sqrt{-g} \,(R-2\Lambda-\alpha C^2) + S_{\rm mat}\,;
\end{equation}
%
$\Mpl^2=1/(8\pi G_N)$, $\Lambda$ is the cosmological constant, $S_{\rm mat}$ is the action of matter, and 
the quadratic Weyl term
in four dimensions reads
\begin{equation}
    C^2=2R^{\mu}{}_{\nu}R^{\nu}{}_{\mu}-\frac23\,R^2 + \mathcal{G}\,,\qquad
\end{equation}  
where
    $\mathcal{G}=R_{\alpha\beta\mu\nu}R^{\alpha\beta\mu\nu}-4R^{\mu}{}_{\nu}R^{\nu}{}_{\mu}+R^2$
is the Gauss-Bonnet combination.\footnote{{In four dimensions, this term can be omitted, as it is a total derivative, and does not influence the equations of motion for the metric field.}}
Since $C^2$ vanishes on an FLRW manifold \cite{Lovelock:1971yv}, the equations of motion for the background will be identical to the ones of $\Lambda$CDM. In what follows, we study the propagation of the degrees of freedom in an expanding universe, where the matter fields are dominated by a single matter component. We consider late-time propagation and assume the matter Lagrangian term in the action $S_{\rm mat}$ to describe a pressureless fluid. We aim to derive constraints on the free parameter $\alpha$ by considering the evolution of the propagating degrees of freedom, 
and eliminate regions of parameter space that give rise to pathological, or patently unphysical behavior.

\section{Dynamics of the Background} \label{Dynamics of the Background}
We consider a spatially flat FLRW background
\begin{equation}
    ds^2=-N^2(t)dt^2+a^2(t)\delta_{ij}dx^i dx^j~,
\end{equation}
where $N(t)$ is the Lapse function, $a(t)$ is the scale factor, both 
functions of the cosmological time coordinate and $\delta_{ij}$ is the Kroenecker delta. The Lagrangian density evaluated on the background in the ADM formalism \cite{Arnowitt:1962hi} reduces to,
%
%
\begin{align}
    \mathcal{L}&=-a^{3} N \Lambda  \,\Mpl^2-a^{3} N \rho \! \left(n\right)-a^{3} n \,\dot\phi 
    -\frac{3 a \,\Mpl^2 \dot{a}^{2}}{N}\,,
\end{align}
where $n$ is the number density of the fluid and $\phi$ is a field imposing number conservation. Applying a variational principle of the action corresponding to this Lagrangian density $\int dt \mathcal{L}$ with respect to $N$, $\phi$, $n$, and $a$, we arrive at the background equations of motion
\begin{align}
    3\Mpl^2H^2&=\Mpl^2\Lambda+\rho\,,\\
    \frac{d}{dt}(na^3)&=0\,,\\
    \phi&=-\int^t N\frac{\partial\rho}{\partial n}\,dt\,,\\
    2\Mpl^2\,\frac{\dot{H}}{N}&=-n\,\frac{\partial\rho}{\partial n}\,.
\end{align}
where $H=\dot{a}/(aN)$ is the Hubble factor. Since for a fluid $p=n\,\frac{\partial\rho}{\partial n}-\rho$, then the last equation can be rewritten as
\begin{equation}
    2\Mpl^2\,\frac{\dot{H}}{N}=-(\rho+p)\,.
\end{equation}
We find that the background equations of motion reduce to the ones of $\Lambda$CDM.\footnote{The continuity equation is merely a consequence of the equations of motion, as
%
    $\dot\rho = \frac{\partial\rho}{\partial n}\,\dot{n} = \rho_{,n}\,\dot{n}=
    \frac{\rho+p}n\,(-3NHn)=-3NH(\rho+p)$.}
We now introduce the following field redefinitions, which we will use throughout the paper to describe the evolution of the background. We define $K = k / (a H)$, $k$ in this context is merely a constant, but it will later on denote the modulus of the wavevector of a propagating mode, $Y = H_0^2 / H^2$, and $\Omega = \rho / (3 \Mpl^2 H^2)$. These redefinitions lead to the following system of ordinary differential equations (ODEs),
\begin{align}
    \frac{K'}K&=-1 +\frac32\,\Omega(1+w)\,,\\
    Y'&=3(1+w)Y\Omega\,,\\
    \Omega'&=-3(1+w)\Omega(1-\Omega)\,,
\end{align}
where $w = p / \rho$ (which is generally a function of cosmic time), and a prime denotes differentiation with respect to the e-fold variable $\mathcal{N} = \ln(a / a_0)$. We impose the following initial conditions: $K(0) \in [10, 100]$, $Y(0) = 1$, and $\Omega(0) \simeq 0.3$. These last conditions reflect the fact that the perturbations are on sub-horizon scales but still inside the linear regime, set the current value of the Hubble rate, and the fraction of the universe's energy density corresponding to non-relativistic matter, respectively.

\section{Metric perturbations} \label{Metric perturbations}

We discuss the metric perturbations in an expanding FLRW background and the selection of an appropriate gauge for studying the propagation of modes. The perturbations are decomposed using an ADM foliation of spacetime \cite{Arnowitt:1962hi}, which leads to the following general line element,
\begin{align}
    ds^2 &= -(1+2\tilde\alpha)\,N^2\,dt^2+
    2 N(\partial_i\chi + a G_i)\,dt\,dx^i \nonumber \\
    &+[(1+2\zeta)a^2\delta_{ij}+2\partial_{i}\partial_j E
    +2a\partial_{(i} C_{j)}+h_{ij}] dx^i dx^j .
\end{align}
%
The four-dimensional diffeomorphism invariance of the system provides the freedom to perform coordinate transformations that simplify the line element. While there are several possible choices of gauge, we opt to introduce and study later on only gauge-invariant fields to transition between them. At this point, in particular, we choose a gauge in which the fields $\zeta$, $E$, and $C_i$ vanish, leading to the simplified line element;
\begin{align}
    ds^2 = -(1+2\tilde\alpha)\,N^2\,dt^2+
    2 N(\partial_i\chi + a G_i)\,dt\,dx^i
    +h_{ij} \,dx^i\,dx^j\,.
\end{align}
To study the propagation of the modes, we focus on the equations of motion for the linear perturbations. Due to the symmetries of the background, the modes decouple from one another, allowing us to simplify the analysis while maintaining generality. In particular, we can consider modes that propagate along the $x$-direction, or equivalently, rotate the coordinate axes to align the $x$-direction with the direction of propagation. This reduces the line element to
\begin{align}
    ds^2 &= -[1+2\tilde\alpha(t,x)]\,N^2\,dt^2+
    2 N\partial_x\chi(t,x)\,dt\,dx  
    + a G_y(t,x)\,dt\,dy+ a G_z(t,x)\,dt\,dz
    \nonumber\\ 
    &+\frac{\sqrt{2}}2\,h_+(t,x)\,(dy^2-dz^2)+\sqrt{2}\,h_\times(t,x)\,dydz\,,
\end{align}
From this, we can extract the Lagrangian density for the gravity sector. For the matter fluid, we introduce its Lagrangian using the simplest form of the Schutz-Sorkin action \cite{Schutz:1977df}.
\begin{equation}
    S_m=-\int d^4x\,\sqrt{-g} [\rho(n)+J^\mu\partial_\mu\phi]\,
\end{equation}
where $n \equiv \sqrt{-J^\mu J_\mu} = \sqrt{ - J^\mu J^\nu g_{\mu \nu}}$ represents the number density of the matter fluid. The four-velocity of the fluid is then given by $u^\mu \equiv J^\mu / n$. The field $\phi$ enforces number conservation, i.e., $\nabla_\mu (n u^\mu) = 0$. Expanding $J^\mu$ and $\phi$ into background and perturbation contributions ($\phi=\phi(t)+\delta\phi$), we define 
\begin{equation}
    J^\mu\,
    \frac{\partial}{\partial x^\mu} 
    = \frac{J^0(t)}{N(t)}\,(1+\delta J^0)
    \,\frac{\partial}{\partial t}
    +\frac{1}{a^2}\delta^{ij}(\partial_j \delta J) \frac{\partial}{\partial x^i}+\frac{1}{a}\,J_V^i \frac{\partial}{\partial x^i}\,,
\end{equation}
where $\partial_i J_V^i = 0$, and with $J^\mu = (J^0(t) / N, \vec{0})$ at the background level. Finally, we can use time reparametrization invariance to set $N(t) = 1$ on the background. With this gauge choice, the propagation of the modes is now uniquely fixed, allowing us to discuss their dynamics in a clear and unambiguous manner.

\section{Propagation of the vector modes} \label{Propagation of the vector modes}

\begin{figure}[t]
    \centering
    \includegraphics[width=13truecm]{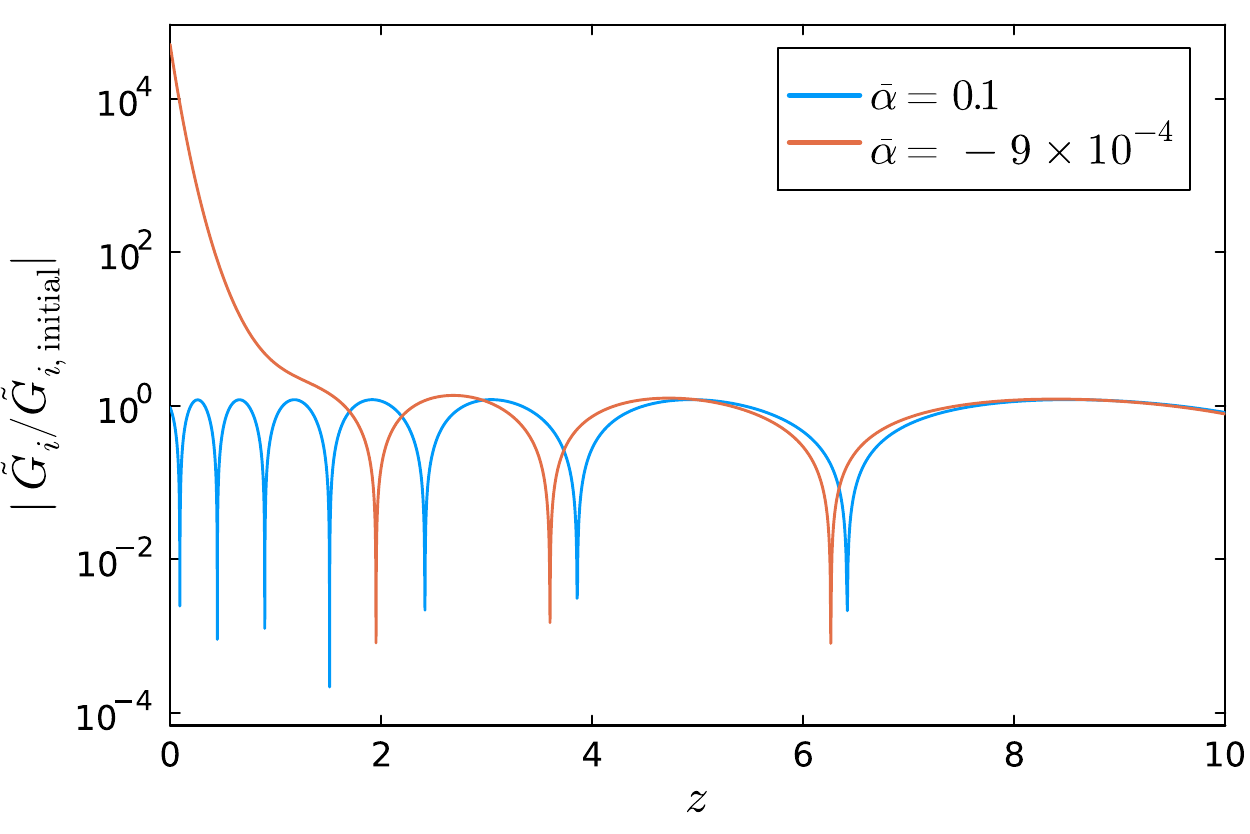}
    \caption{Stability and instability of the gauge invariant vector modes as a function of the redshift. The stability is obtained for $\bar\alpha>0$. In the top panel, we have chosen $\bar\alpha=0.1$, whereas in the lower one, $\bar\alpha=-9\times10^{-4}$. We observe the instability of the mode in the limit of the present time for the negative choice of $\bar{\alpha}$. In both cases $K_0=k/(a_0 H_0)=10$. We have assumed the matter fields to be a pressureless fluid, that is $c_s^2=0=\eta=0=w$. To obtain the figure, we first integrated the background equations of motion by setting $\Omega(z=0)=0.3$, $Y(z=0)=1$. For the perturbations, we have chosen instead, as initial conditions, ${\tilde G}_i=10^{-5}=-{\tilde G}_i'$ at $\mathcal{N}=-5$.}
    \label{fig:Vector_Stability_Instability}
\end{figure}

As already discussed, we have the freedom to define gauge-invariant fields to describe the propagating vector degrees of freedom, as follows, 
\begin{equation}
    J_V^i=\tilde{J}_V^i-a\,\frac{J^0(t)}{N(t)}\,\delta^{ij}\partial_t(C_j/a)\,, \\
    G_i = \tilde{G}_i + \frac{a}{N}\,\partial_t(C_i/a)\,.
\end{equation}
By making these field redefinitions, the fields $C_i$ become gauge modes and entirely vanish from the action. The fields $\tilde{J}_V^i$ act as Lagrange multipliers, which can be integrated out by solving their algebraic equations of motion:
\begin{equation}
    {\tilde J}_V^i=-n(t)\,\delta^{ij}\,\tilde{G}_j\,.
\end{equation}
Next, we introduce a Fourier decomposition for the remaining vector components and set $N(t) = 1$. The equation of motion for these components is 
\begin{equation}
    \ddot{\tilde{G}}_i + H \dot{\tilde{G}}_i + \left[\frac{k^2}{a^2 H^2} + \frac{1}{2\bar\alpha}\frac{H_0^2}{H^2}  \right] H^2 \tilde{G}_i =0\,,\label{eq:mass_V}
\end{equation}
where $\bar \alpha/ H_0^2 =\alpha$, and assumed $H\neq0$. Using the e-fold variable $\mathcal{N} = \ln(a/a_0)$ as the independent time variable, we rewrite the dynamical equation as
\begin{equation}
{\tilde G}_i''+\left[ 1-\frac{3}{2}(1+w)\,\Omega \right]    {\tilde G}_i' + \left( K^2 +\frac{Y}{2\bar\alpha} \right) {\tilde G}_i=0.
\end{equation}
To ensure the stability of the new propagating modes $\tilde{G}_i$ during their evolution, we impose $\alpha > 0$. Under this constraint, the modes will remain stable at all scales. In the limit where $\bar{\alpha} \to 0$, we recover the GR limit, with $\tilde{G}_i \to 0$. Two examples of stability and instability behaviors for the gauge-invariant vector modes as a function of redshift are illustrated in Fig.~(\ref{fig:Vector_Stability_Instability}).
It is important to note that strong constraints apply only to sub-horizon modes, because not affected by cosmic variance. From Eq.~\eqref{eq:mass_V}, it is clear that the modes have an effective squared-mass proportional to  $1/(2\alpha)$. This implies that for $\alpha>0$ the modes propagate without any instability. However, from the Lagrangian density 
\begin{equation}
    \frac{\mathcal{L}}{\Mpl^2} = -\frac{a\alpha k^2}{2}\,\dot{\tilde{G}}_i\dot{\tilde{G}}_j\delta^{ij}
    +\left[ \frac{\alpha k^4}{2a} + \frac{ak^2}{4} \right] \tilde{G}_i \tilde{G}_j\delta^{ij}\,,
\end{equation}
one notes that 
for $\alpha >0$, the kinetic term is negative. Here, as we have already stated in the introduction, we are in the presence of a classically stable ghost mode.
Then to avoid vacuum quantum instabilities and to keep the theory unitary one needs to follow the ``fakeon'' prescription introduced in \cite{Anselmi:2018kgz}.

\section{Propagation of the tensor modes} \label{Propagation of the tensor modes}

We now turn our attention to the sector of gravitational waves, which correspond to the degrees of freedom associated with the tensor components $h_{ij}$. These components are intrinsically gauge invariant. The tensor perturbations $h_{ij}$ can be expressed in terms of the two polarizations as,
\begin{equation}
    h_{ij} = h_{+}\epsilon_{ij}^{+} + h_{\times}\epsilon_{ij}^{\times}\,,
\end{equation}
where, for each polarization, the following conditions hold: $\delta^{ij} \epsilon_{ij} = 0$, $\delta^{ik} \partial_k h_{ij} = 0$, $\epsilon_{ij} = \epsilon_{ji}$, $\epsilon_{ij} \epsilon_{ji} = 1$, and $\epsilon_{ij}^{+} \epsilon_{ji}^{\times} = 0$.
By expanding the action in the tensor perturbations, we derive the following lagrangian density for these modes,

\begin{align}
    \frac{\mathcal{L}}{\Mpl^2}&=\sum_{\lambda=+,\times}
    \left[-\frac{a^{3} \alpha \ddot{h}_\lambda^{2}}{4}+\left(\frac{\alpha  \,k^{2} a}{2}-\frac{a^{3} H^{2} \left(3 \Omega  w +3 \Omega -4\right) \alpha}{8}+\frac{a^{3}}{8}\right) \dot{h}_\lambda^{2}+\left(-\frac{\alpha  \,k^{4}}{4 a}-\frac{a \,k^{2}}{8}\right) h_{\lambda}^{2}\right].
\end{align}
This leads to the following equations of motion for the gravitational waves,
%
\begin{align}
    \ddddot{h}_{\lambda}&=-6 H \dddot{h}_{\lambda}-\left(\frac{2 k^{2}}{a^{2}}+H^{2} \left(11-6 \Omega  \left(1+w \right)\right)+\frac{1}{2 \alpha}\right) \ddot{h}_{\lambda}\nonumber\\
    &-\left(\frac{2 k^{2}}{a^{2}}+\frac{9 H^{2} \left(\frac{4}{3}+\left(1+w \right) \left(c_{s}^{2}-\frac{4}{3}\right) \Omega \right)}{2}+\frac{3}{2 \alpha}\right) H\dot{h}_{\lambda}-\left(\frac{k^{2}}{a^{2}}+\frac{1}{2\alpha}\right) \frac{k^2}{a^2}\, h_{\lambda}
\,.\label{eq:GW_full_N}
\end{align}
In the limit $\alpha \to 0$, the equation of motion simplifies to the following, 
\begin{align}
    \ddot{h}_\lambda+3H\dot{h}_\lambda+\frac{k^2}{a^2}h_\lambda=0\,\label{eq:hGR}
\end{align}
that is recovering the standard equation for tensor perturbations in GR on an expanding background. Next, we find it convenient to define two time-dependent quantities $\eta \equiv n^2 \rho_{,nnn}/\rho_{,n}$, $c_s^2\equiv n\rho_{,nn}/\rho_{,n}$, and $K = k/(aH)$. We obtain the following transformed equation of motion for the gravitational waves,

\begin{align}
    &h''''_\lambda+\left[6-9\left(1+w\right) \Omega \right] h'''_\lambda+\left(2 K^{2}+11+\frac{27 \Omega^{2} \left(1+w \right)^{2}}{4}+18 \left(1+w \right) \left(c_{s}^{2}-\frac{5}{6}\right) \Omega +\frac{Y}{2 \bar\alpha}\right) h''_\lambda\nonumber\\
    &+\left([2-3(1+ w) \Omega] K^{2}+6-\frac{27 \left(1+w \right)^{2} \left(c_{s}^{2}-\frac{1}{3}\right) \Omega^{2}}{4}-9(1+w ) \Omega\! \left(c_{s}^{2}+\frac{3 \eta}{2}+1\right) -\frac{3 Y \left(\Omega  w +\Omega -2\right)}{4 \bar\alpha}\right) h'_\lambda\nonumber\\
    &+\left(K^{4}+\frac{Y \,K^{2}}{2 \bar\alpha}\right) h_\lambda = 0\,,\label{eq:GW_dyn}
\end{align}
where, as for the vector modes, a prime denotes differentiation with respect to the e-fold variable. This is a fourth-order differential equation, which implies that, compared to GR, we expect two additional initial conditions for each of the two polarizations. In the limit $\alpha \to 0$, we can find an analytical solution by perturbatively expanding $h_\lambda$ in powers of the dimensionless parameter $\alpha H_0^2$.
\begin{equation}
    h_\lambda = h_\lambda^{\rm GR}+\alpha H_0^2 \,h_\lambda^{(1)}+ \mathcal{O}(\alpha H^2_0)^2,
\end{equation}
where $h_\lambda^{\rm GR}$ satisfies Eq.\ \eqref{eq:hGR}, namely
\begin{equation}
    {h_\lambda^{\rm GR}}''
    +3\left(1-\tfrac12 (1+w)\Omega\right){h_\lambda^{\rm GR}}'+K^2h_\lambda^{\rm GR}=0\,.
\end{equation}
Order by order in $\alpha H_0^2$, one can construct the solution by knowing the solution at the previous order. For example,
%
\begin{align}
    {h}^{(1)}_\lambda{}''
    &+3\left(1-\tfrac12 (1+w)\Omega\right) {h}^{(1)}_\lambda{}'+K^2 {h}^{(1)}_\lambda \nonumber \\
    &=6\,\frac{H^2}{H_0^2}(1+w)(5+3c_s^2)\Omega {h_\lambda^{\rm GR}}'
+12\,\frac{H^2}{H_0^2}(1+w)\Omega\,K^2h_\lambda^{\rm GR}\,.\label{eq:order_by_order_GR_GW}
\end{align}
%
We observe that the source term for $h_\lambda^{(1)}$ in the right-hand side becomes stronger at higher redshifts, and the homogeneous solution of the l.h.s.\ of Eq.~\eqref{eq:order_by_order_GR_GW} can be re-absorbed into the original GR contribution.
For general $\alpha$ we enter the full modified gravity regime. In the subhorizon limit, we can approximate the equations of motion as
\begin{align}
    h''''_\lambda+\left[6-9\left(1+w\right) \Omega \right] h'''_\lambda+2 K^{2} h''_\lambda+[2-3(1+ w) \Omega] K^{2} h'_\lambda+K^{4} h_\lambda \approx 0\,.
\end{align}
In the Wentzel-Kramer-Brillouin (WKB) approximation, the modes would propagate with the speed of light, since $h''''_\lambda\approx -K^{4} h_\lambda$. Therefore, the constraints for the speed of propagation of the gravitational waves compared to the one of electromagnetic waves, are then satisfied.
\begin{figure}[t]
    \centering
    \includegraphics[width=13truecm]{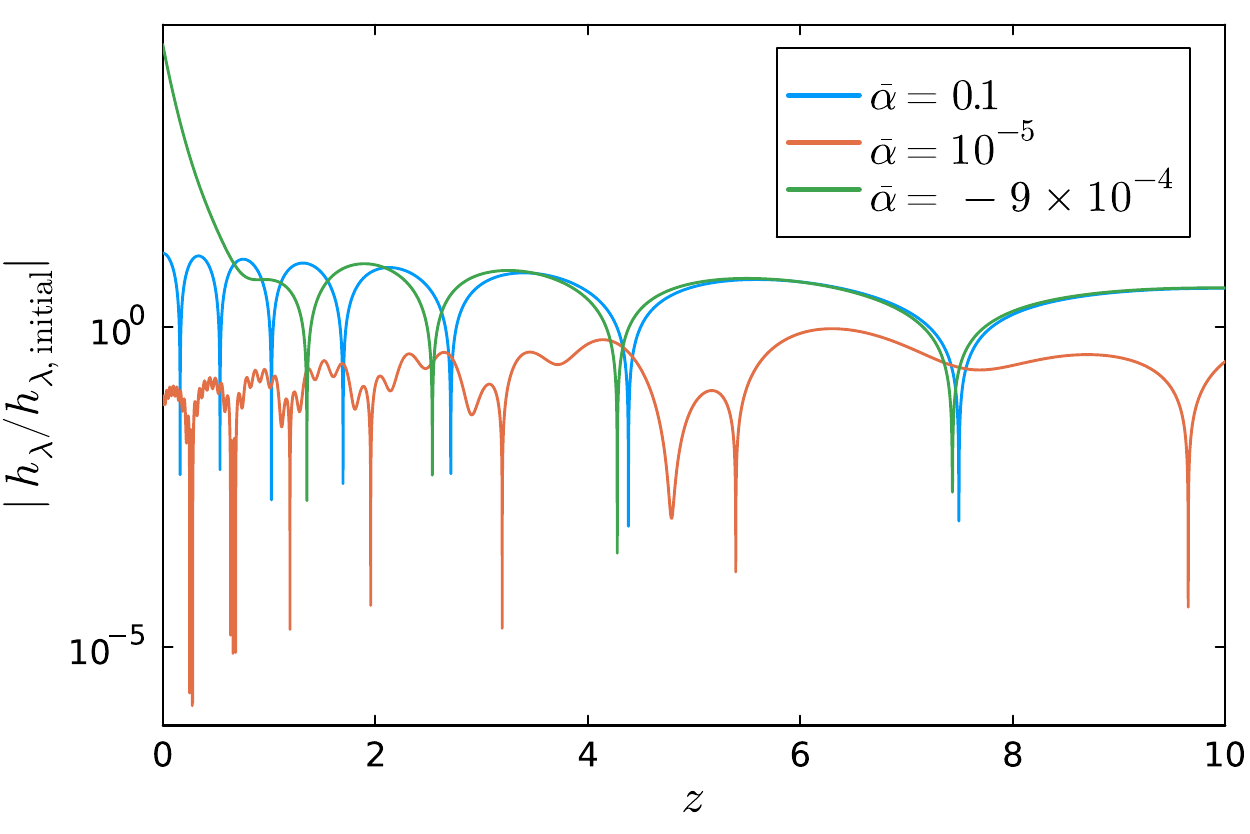}
    \caption{Stability and instability of the two gravitational waves solving Eq.~\eqref{eq:GW_full_N}. This instability is present for negative $\alpha$ where we have fixed $\bar{\alpha}=-9\times10^{-4}$. Once more, we have set $k/(a_0 H_0)=10$, and the same initial conditions for the background as in Fig.~\ref{fig:Vector_Stability_Instability}. To generate this figure, we have fixed $h_\lambda=10^{-5}(1+i)$, 
 $h'_\lambda=0.4h_\lambda$, $h''_\lambda=0.01h_\lambda$, $h'''_\lambda=-0.02h_\lambda$ at $\mathcal{N}=-5$. If $h_{\lambda}$ is a solution, then $ C\,h_{\lambda}$ is also a solution, $C$ being an arbitrary constant. Therefore, we plot the physically relevant quantity related to the growth, which is $h_\lambda / h_{\lambda,\mathrm{initial}}$.}
    \label{fig:GW_Tensor_Instability}
\end{figure}
Similar to the case of the vector degrees of freedom, we observe in Fig.~\ref{fig:GW_Tensor_Instability} that we require $\alpha > 0$ in the tensor sector to eliminate instabilities in the late time limit of the modes' propagation. To generate Fig.~\ref{fig:GW_Tensor_Instability}, we have integrated Eq.\ \eqref{eq:GW_dyn} setting the background fluid to be pressureless, so that $\eta=0=c_s^2=0=w$.

\section{Propagation of the scalar modes} \label{Propagation of the scalar modes}

So far we have found that stability of the propagation for the modes is ensured when $\alpha>0$, and in any case the gravitational waves in this theory propagate with the speed of light. Sufficiently small values of $\alpha H_0$ are enough to suppress the propagation of the vector modes present in the theory. However, we should expect an impact on cosmological observables coming from the dynamics of the scalar perturbations, in particular affecting the growth of perturbations. Therefore, we now proceed to discuss the propagation of the scalar degrees of freedom. To begin, we introduce the following variable
\begin{equation}
    \delta J^0 = \frac{\rho}{n\rho_{,n}}\,\delta-\tilde\alpha\,,
\end{equation}
where $\delta=\delta\rho/\rho$. Next, we define the gauge-invariant combinations of fields, $\Psi$, $\Phi$, and $\tilde{\delta}_{\rm m}$, as follows, 

\begin{align}
    \tilde\alpha &= 
\Psi -\dot\chi+2 a^{2} H\, \frac{{\rm d}}{{\rm d} t}\!\left(\frac{E}{a^{2}}\right) +a^{2} \frac{{\rm d^2}}{{\rm d} t^2}\!\left(\frac{E}{a^{2}}\right),\\
\zeta &= 
-\Phi -H \chi +a^{2} H\, \frac{{\rm d}}{{\rm d} t}\!\left(\frac{E}{a^{2}}\right),\\ 
\delta &= 
{\tilde\delta}_{\rm m}+3 (1+w ) H \chi -3 (1+w ) a^{2} H\, \frac{{\rm d}}{{\rm d} t}\!\left(\frac{E}{a^{2}}\right),
\end{align}
where we have set $N(t) = 1$ so that $t$ represents the cosmic time, and $w = p / \rho$ is the possibly time-dependent equation of state parameter. 
With these field redefinitions in place, we find that $\chi$ and $E$ are pure gauge fields, i.e.\ they automatically disappear from the dynamical equations of motion. To simplify the equations of motion, we find it convenient to perform the following field redefinition,

\begin{equation}
    \tilde\delta_{\mathrm{m}} = \delta_{\mathrm{m}}+\frac{3n\rho_{,n}}{\rho}\,\Phi\,,
\end{equation}
so that $\delta_{\mathrm{m}}$ is also gauge invariant, being a linear combination of gauge-invariant variables. We can integrate out all fields except for $\delta_{\rm m}$ and $\Phi$---showing the existence in this theory of an extra scalar degree of freedom compared to GR---which are required to satisfy a system of 2nd order ODE. We explicitly write down these equations in the Appendix, for the convenience of the reader.

To solve numerically these equations, and to have also a better analytical understanding of the behavior of the solutions, we find it convenient to transform the system of Eqs.~\eqref{fullDelta} and~\eqref{fullPhi}, into a single fourth-order ODE for the field $\delta_m$. To reach this goal, we begin by formally solving the equation of motion for $\delta_m''$, Eq.~\eqref{fullDelta}, in terms of $\Phi'$, i.e., $\Phi' = Z_1(\Phi, \delta_m'', \delta_m', \delta_m)$. We then substitute this expression for $\Phi'$ into the equation of motion for $\Phi$, Eq.~\eqref{fullPhi}, which can be formally solved for $\Phi$ in terms of $\delta_m$ and its time derivatives, as $\Phi = Z_0(\delta_m''', \delta_m'', \delta_m', \delta_m)$. For consistency, we require that $Z_0' = Z_1(\Phi = Z_0, \delta_m'', \delta_m', \delta_m)$. This leads to a fourth-order ordinary differential equation (ODE) for $\delta_m$, and it is this equation that we numerically solve. For this work, from now on, we focus on the case that is of interest to us, where matter is modeled by a dust fluid.\footnote{We aim to give an upper bound on the value of $\alpha H_0^2$ by studying the influence of this parameter on the matter power spectrum, which is dominated by a cold dust component.} Furthermore, to have a solution of the equations for small values of $|\alpha| H_0^2\ll1$ (but still considering $|\alpha|^{-1}\ll M_{\rm cutoff}^2$) which cannot be easily found numerically, it is necessary to have a perturbative solution built out of the GR solution. For this aim, we perturbatively expand $\delta_m$ in powers of $\alpha H_0^2$,
\begin{equation}
    \delta_m=\delta_m^{\rm GR}+ \alpha H_0^2 \delta_m^{(1)}+\dots
\end{equation}\label{Perturbative Expansion}
and solve the fourth-order equation for $\delta_m$ order by order in $\alpha H_0^2$.\footnote{The theory, when interpreted as a low energy effective theory, will have a cutoff scale, $M_{\rm cutoff}$, which we suppose not known a priori. In this case, the theory we are now discussing makes sense as a low energy effective theory if $1/|\alpha|\ll M_{\rm cutoff}^2$, or when $1/|\bar\alpha|\ll M_{\rm cutoff}^2/H_0^2$. For values of $|\bar\alpha|$ not satisfying this inequality, the new propagating modes will acquire a mass above the cutoff scale and can be integrated out from the theory. Therefore, for this iterative expansion, we suppose that $1\ll 1/|\bar\alpha|\ll M_{\rm cutoff}^2/H_0^2 $.} At the lowest order, we find the GR equation for the matter growing mode in the $\Lambda$CDM background:
\begin{align}
    {\delta_m^{\rm GR}}''+\frac{K^2(8-6\Omega)+27\Omega(2-\Omega)}{4K^2+18\Omega}\, {\delta_m^{\rm GR}}'-\frac{3\Omega K^2}{2K^2+9\Omega}\, {\delta_m^{\rm GR}}=0\,, \label{Full GR Equation}
\end{align}
which, in the subhorizon limit, simplifies to
\begin{equation}
    {\delta_m^{\rm GR}}''+\frac{(4-3\Omega)}{2}\, {\delta_m^{\rm GR}}'-\frac{3\Omega}{2}\, {\delta_m^{\rm GR}}=0\,.
\end{equation}
In the deep matter era, where $\Omega = 1$, the usual growing mode solution is recovered: $\delta_m \propto a$. Expanding the quantity $Z_0$ to lowest order in $\alpha$, we find the standard Poisson equation,
\begin{align}
    \Phi_{\rm GR} &= \frac{9\Omega}{K^2(2K^2+9\Omega)}\,{\delta_m^{\rm GR}}'-\frac{3\Omega}{2K^2+9\Omega}\,{\delta_m^{\rm GR}}\nonumber\\
    &\approx -\frac{3\Omega}{2K^2}\,{\delta_m^{\rm GR}}\,,
\end{align}
where in the second line, we assume that ${\delta_m^{\rm GR}}' \approx {\delta_m^{\rm GR}}$ and that the mode propagates on sub-horizon scales. At the next order in $\alpha H_0$, we find the following equation for $\delta_m^{(1)}$,
\begin{align}
    {\delta_m^{(1)}}''&+\frac{K^2(8-6\Omega)+27\Omega(2-\Omega)}{4K^2+18\Omega}\, {\delta_m^{(1)}}'-\frac{3\Omega K^2}{2K^2+9\Omega}\, {\delta_m^{(1)}}=S_0\,,\label{PerturbativeAnsatzLHS}
\end{align}
where
\begin{align}
    S_0&=\frac{16\! \left[\frac{1215 \Omega^{2} {\delta_m^{\rm GR}}'}{8} -K^{6} \delta_m^{\rm GR} -\left(\frac{27}{4} \Omega  \delta_m^{\rm GR} +\frac{3}{2} \delta_m^{\rm GR} -9 {\delta_m^{\rm GR}}'\right)\! K^{4}+\left(\left(\frac{135 \Omega}{2}+\frac{9}{2}\right)\! {\delta_m^{\rm GR}}'-\frac{81 \Omega^{2} \delta_m^{\rm GR}}{4}\right)\! K^{2}\right] \Omega  \,H^{2}}{\left(2 K^{2}+9 \Omega \right)^{2} H_{0}^{2}}\label{PerturbativeAnsatzRHS}\,.
\end{align}
The homogeneous solution is the same as that for the GR operator, and it will be dominated by the growing mode. The non-trivial part consists of the particular solution of Eq.~\eqref{PerturbativeAnsatzLHS}. In the sub-horizon limit, we have
\begin{align}
    S_0&\approx -\frac{4K^2\Omega H^2}{H_0^2}\,\delta_m^{\rm GR}
    \approx -4K_0^2\,(1+z)^2\,\delta_m^{\rm GR}\,,
\end{align}
where the constant $K_0=k/(a_0 H_0)$ satisfies $K_0 \simeq  100$, and we have assumed a propagation in the deep matter era at which $\Omega\approx1$. For consistency, the source $S_0$ should not result in $ |\alpha| H_0^2 |\delta_m^{(1)}| > |\delta_m^{\rm GR}|$. This condition requires
\begin{equation}
|\alpha| H_0^2 S_0\ll |\delta_m^{\rm GR}|\,,\qquad
|\alpha| H_0^2\,(1+z)^2\,K_0^2\ll1\,, 
\label{eq:ineq_1}
\end{equation}
%
during dust domination or as long as the theory remains within the EFT regime, that is for $1/|\bar\alpha|\ll (M_{\rm cutoff}/H_0)^2$. At certain high redshifts for a given value of $\alpha$ inside the EFT regime, the inequalities in \eqref{eq:ineq_1} will be no longer satisfied. In this case, it becomes necessary to study the dynamics of the perturbations outside the GR limit, i.e.\ solving Eqs.~\eqref{fullDelta} and~\eqref{fullPhi}, since the perturbative solution given above cannot be trusted any longer. In this case, if these dynamics do not lead to a consistent phenomenology, the theory as it is cannot be accepted as a viable theory, and some corrections to it, including possibly some high-redshift modifications of the theory, need to be implemented.

\subsection{Pure modified dynamics regime} \label{Pure modified dynamics regime}

\begin{figure}[ht]
    \centering
    \includegraphics[width=13truecm]{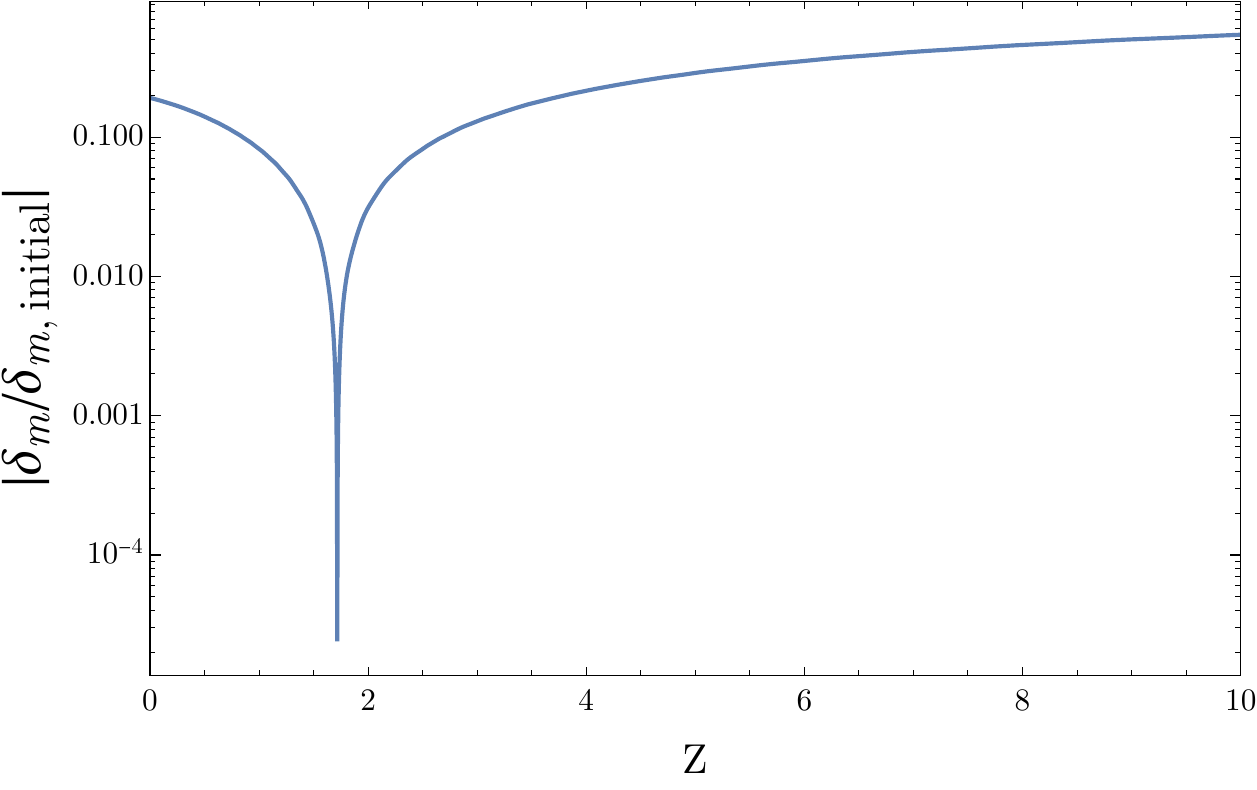}
    \caption{The evolution of the scalar mode in the pure modified regime $\delta_m$ normalized by its initial condition $\delta_{m,\mathrm{initial}}$, as a function of redshift, the solutions are obtained by numerically integrating \eqref{eq:modifiedregime}, where we have taken a sub-horizon approximation. The initial conditions of this mode $\delta_m$ and its derivatives were set to $10^{-5}$ at $\mathcal{N}=-10$, and we have fixed $k/(a_0 H_0)=100$.  
    We observe that this exotic mode does not grow at late times and hence does not recover the formation of structure.} 
    \label{fig:scalarInstability}
\end{figure}

When we do not restrict ourselves to small values of $\alpha H_0^2$, or when the perturbative solution cannot be longer trusted, the closed fourth-order equation of motion for $\delta_m$, in the sub-horizon limit, yields 
\begin{equation}\label{eq:modifiedregime}
    \delta_m''''+9(1-\Omega)\delta_m'''
    +K^2\delta_m''+K^2\bigl(2-\tfrac32\Omega\bigr)\delta_m'+\tfrac12 K^2\Omega\delta_m\approx0\,.
\end{equation}
By considering solutions of the form $\delta_m \propto e^{-i\omega t}$, and assuming the WKB approximation $|\delta'|=H^{-1}|\dot{\delta}|\simeq \frac\omega H |\delta_m|\gg |\delta_m|$, and $\omega\simeq k/a$, 
we find that the mode propagates with the speed of light. Specifically, this is indicated by the approximate dispersion relation $\omega^2 (\omega^2 - k^2/a^2) \approx 0$, where the mode with zero speed of propagation is simply the GR dust mode.\footnote{This mode is not strongly coupled just because, as usual, we have neglected the pressure of the dust component.} This behavior contrasts sharply with the dynamics in GR. It suggests that $\delta_m$ contains a stable propagating component. In this regime, the parameter $\alpha$ no longer plays a role, and it is unclear whether the equation of motion admits an unstable mode.\footnote{For instance, the last term in Eq.~\eqref{eq:modifiedregime} suggests a positive effective mass term for the modes.} More precisely, in this limit, it raises the question: {\sl has the Jeans instability been suppressed?}
To explore this further, we plot the solution of \eqref{eq:modifiedregime}, which is an approximation valid only for modes on sub-horizon scales in Fig.~\ref{fig:scalarInstability}, as a function of redshift. 
We observe that the mode does not exhibit any divergent growth and fails to recover the Jeans instability. This indicates that the mode behaves in a way completely different from the GR case, giving serious phenomenological problems, as it does not exhibit the expected growth associated with structure formation. This behavior is vastly different from the corresponding result in GR.


\subsection{Analytical constraints on $\alpha$ from the scalar sector} \label{Analytical constraints on alpha from the scalar sector}

In the context of these modified cosmologies, we can constrain the parameter space of these models by considering the evolution of scalar perturbations. Specifically, we impose two key requirements: (i) the model must reproduce the growth of scalar matter density perturbations, which serve as the seeds for large-scale structure formation observed in the present universe, and (ii) deviations from GR must be sufficiently small to ensure consistency with current observational constraints. The evolution of scalar matter perturbations $\delta_m$, is governed by the equations of motion given by \eqref{fullDelta} and \eqref{fullPhi}. By appropriately manipulating these equations, one can reduce them to a fourth-order ordinary differential equation (ODE) in terms of $\delta_m$, in the procedure described earlier in this section. The solution of this ODE for varying values of the parameter $\alpha$ allows us to examine the impact of these modifications without approximations on the evolution of perturbations over time and compare with corresponding results from GR. A key feature in the formation of large-scale structure is the presence of
a growing 
regime during the era of matter domination. This transition is critical for the formation of stars, galaxies, and the larger cosmic structures observed today. In the standard cosmological model, this process is driven by the Jeans instability, which amplifies density perturbations and enables their growth. To remain consistent with observations, any modification to GR must permit the Jeans instability to grow according to a power-law, as we expect in $\Lambda$CDM cosmology.

\begin{figure}[!tbp]
  \centering
  \includegraphics[width=13truecm]{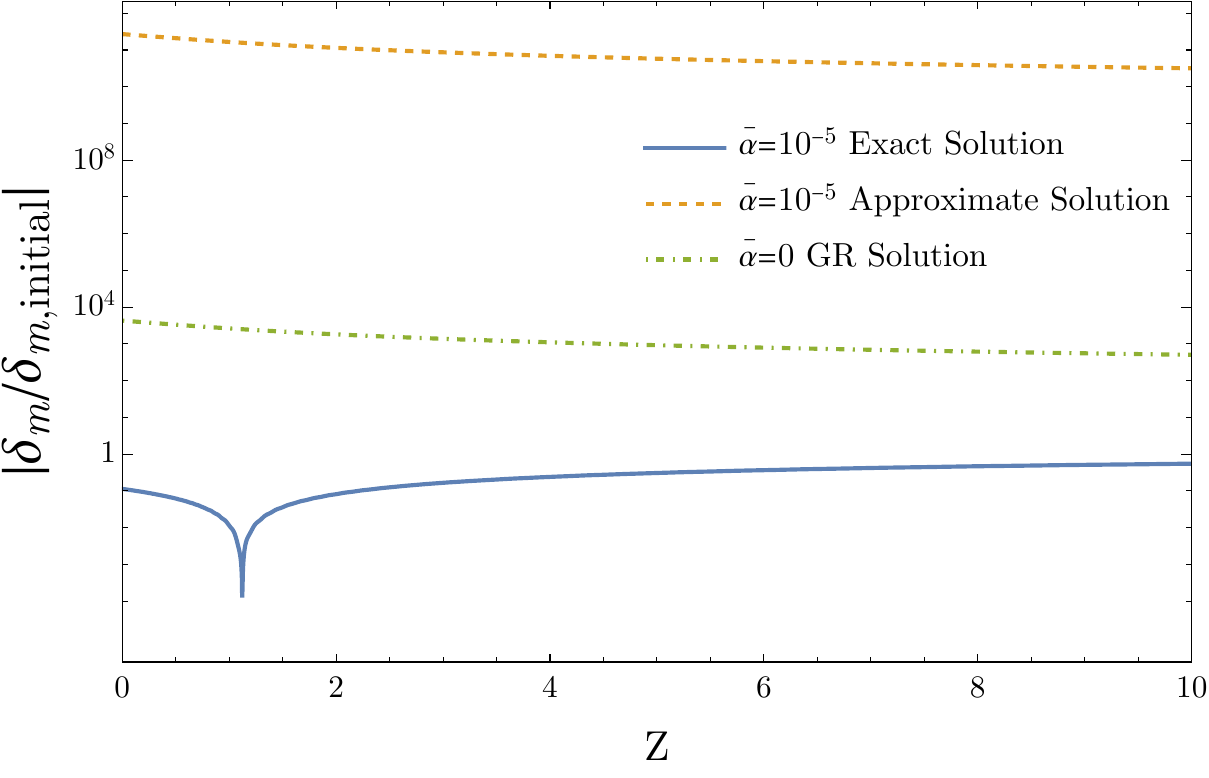}
  \hfill
  \includegraphics[width=13truecm]{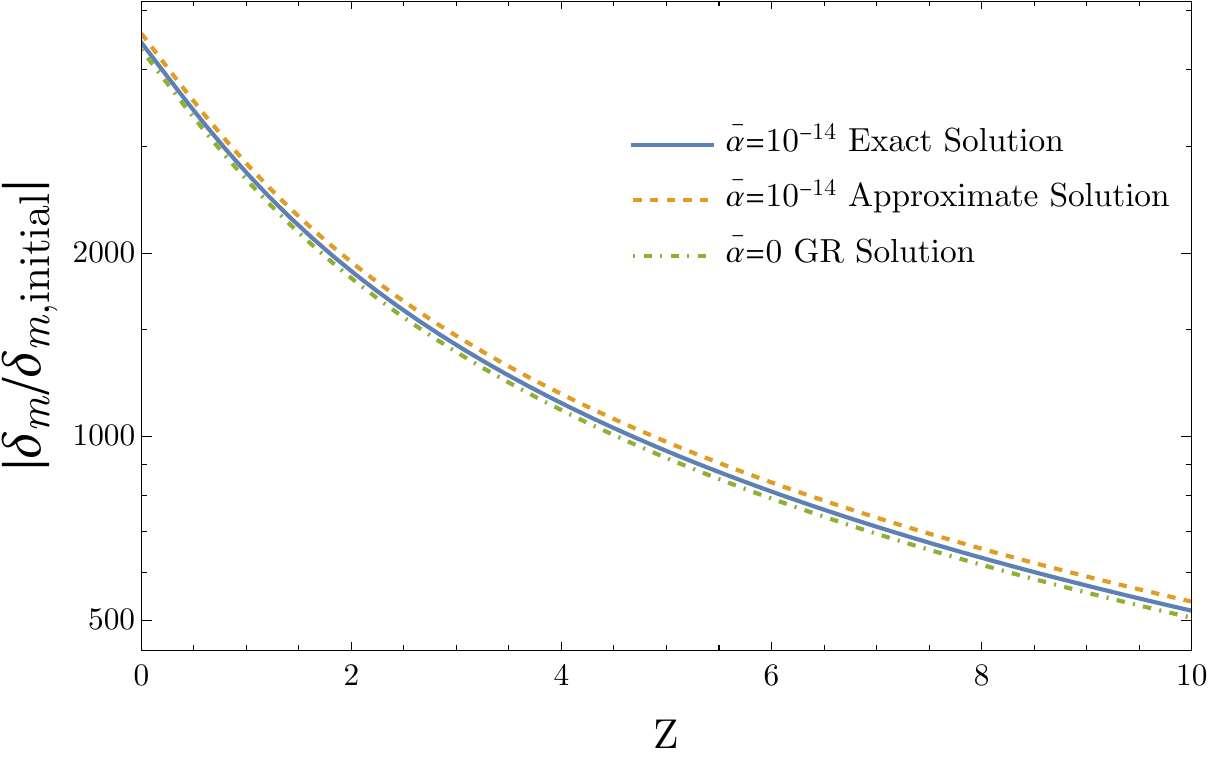}
  \label{fig:f2}
  \caption{The evolution of the scalar perturbation $\delta_m$, normalized by its initial condition $\delta_{m,\mathrm{initial}}$, evolving as a function of redshift, for values of $\Bar{\alpha}=10^{-5}$ and $\Bar{\alpha}=\alpha H_0^2=10^{-14}$ in the top and bottom panels respectively, the exact solutions are obtained by numerically integrating the fourth order ODE for $\delta_m$ described by \eqref{FourthOrderSystem}. The approximate solutions are described by \eqref{Perturbative Expansion}, where the first-order term is obtained from \eqref{PerturbativeAnsatzLHS} and \eqref{PerturbativeAnsatzRHS}. 
  The GR solution is obtained from \eqref{Full GR Equation}. We have fixed $(\frac{k}{a_0 H_0})=100$, and the initial condition for all $\delta_m$, $\delta^{\text{GR}}_m$, $\delta^{(1)}_m$ and their derivatives, in all 3 cases, to $10^{-5}$ at $\mathcal{N}=-10$.} 

\label{fig:K0=10_FourthOrderSystem_ComparisonOfJeansInstabilityForDifferentAlpha}\end{figure}

\begin{figure}[htpb!]
    \centering
    \includegraphics[width=14truecm]{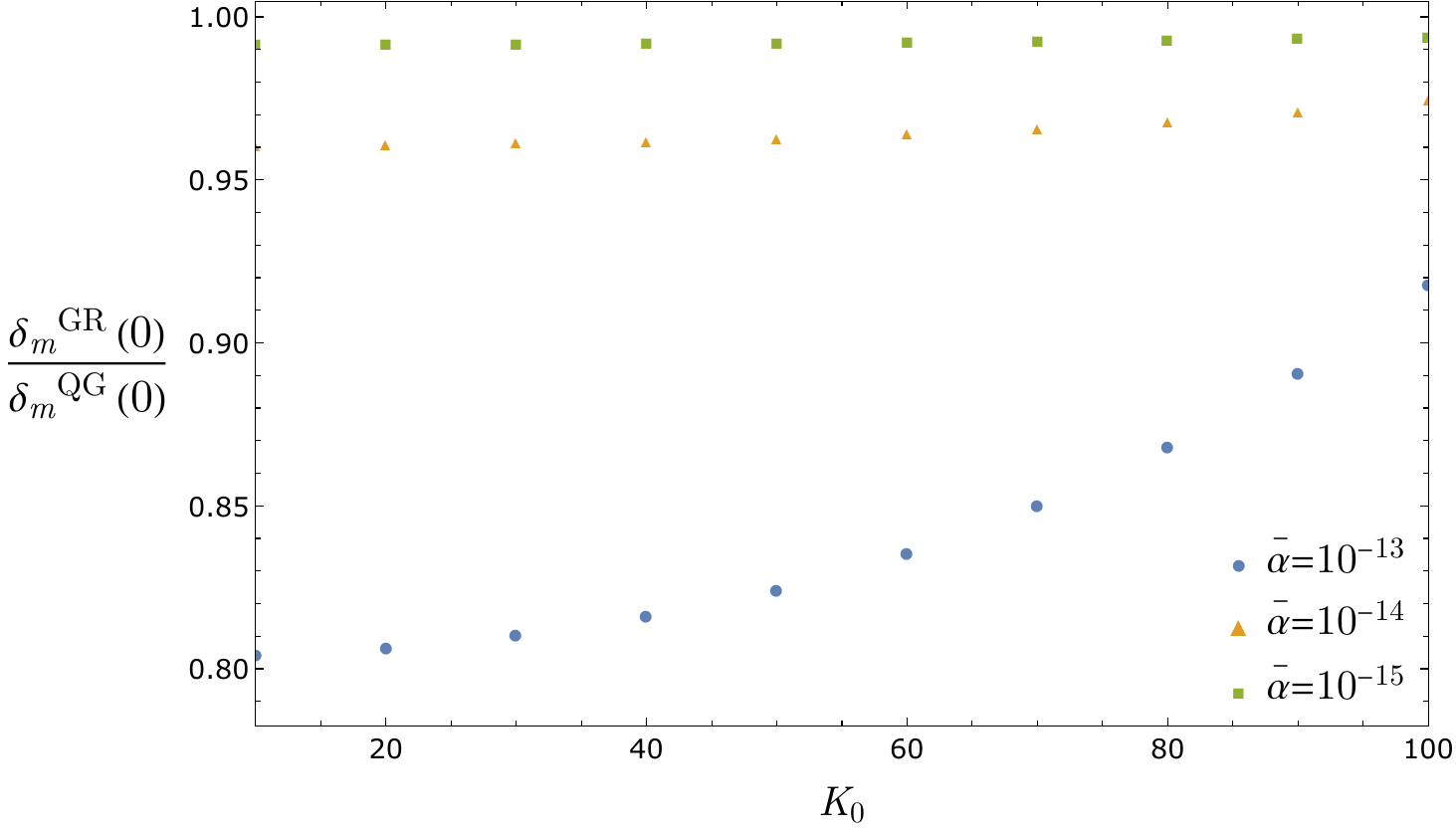}
    \caption{The ratio between the scalar perturbation in quadratic gravity $\delta^{\text{QG}}_m$ and GR $\delta^{\text{GR}}_m$ at the present, for values of $K_0 \equiv k/(a_0 H_0) \in [10,100]$. For $\bar{\alpha}=\{10^{-13},10^{-14}\}$ the system was solved exactly from \eqref{FourthOrderSystem}. Whereas for $\bar{\alpha}=10^{-15}$ $ \delta^{\text{QG}}_m \equiv \delta^{\text{GR}}_m + \bar{\alpha} \delta^{(1)}_m + \mathcal{O} (\bar{\alpha}^2)$ was given by the perturbative approximation \eqref{PerturbativeAnsatzLHS} up to linear order. The initial conditions were fixed as $\delta^{\text{GR} } _m$ and $\delta^{(1)} _m$ and their derivatives to $10^{-5}$ at $\mathcal{N}=-10$.} 
    \label{fig:ListPlotPointsSmallAlpha}
\end{figure}

As shown in the bottom panel FIG.~\ref{fig:K0=10_FourthOrderSystem_ComparisonOfJeansInstabilityForDifferentAlpha}, for sufficiently large values of
$\alpha$, the Jeans instability is suppressed, preventing the perturbations from growing. This suppression signifies a failure to recover the observed growth of the structure, placing a strong constraint on the permissible values of $\alpha$. In addition to this, we observe that for $\bar{\alpha}$ too large in magnitude, the perturbative approximation described by \eqref{PerturbativeAnsatzLHS} fails to describe the system accurately. For smaller $\bar{\alpha}$ we recover the growth of perturbations, and the accuracy of the approximate solution to \eqref{PerturbativeAnsatzLHS}. We observe this for $\bar{\alpha} = 10^{-14}$ in the bottom panel of FIG.~\ref{fig:K0=10_FourthOrderSystem_ComparisonOfJeansInstabilityForDifferentAlpha}. By requiring that the model recovers the observed growth of structure and that deviations from GR are small, we can place stringent constraints on the parameter $\alpha$ and, by extension, the parameter space of the quadratic gravitational model we are considering. These constraints are essential for evaluating the viability of any proposed modification to GR, ensuring that such models remain consistent with both observational data and theoretical expectations.
However, this line of reasoning presents a challenge. The constraints on the parameter $\alpha$ derived from the evolution of perturbations are sensitive to the choice of the initial condition for the ratio $k/(a_0 H_0)$, inversely related to the typical length scale of the perturbation, whose range lies within $[10,100]$, and as a result, the derived constraints on $\alpha$ can vary considerably depending on where we select a point within this range. We are restricted to choosing values of the initial condition $K_0=k/(a_0 H_0) \in [10,100]$ because, for $K_0 < 10$, the modes will be dominated by cosmic variance, and for $K_0 > 100$ we enter the non-linear regime of cosmological perturbation theory. This sensitivity to the choice of $k$, the scale for the perturbations, introduces a degree of uncertainty into the results, making it difficult to derive robust, universally valid constraints on $\alpha$ based on a single choice of $k/(a_0 H_0)$. To address this issue and derive constraints on $\alpha$ that are independent of the specific choice of the wavenumber, we propose a more comprehensive approach. We solve the system for a range of values spanning the full possible range of these initial conditions for $k/(a_0H_0)$. By sampling a range of values within the interval $[10,100]$, we can ensure that the constraints on $\alpha$ are valid across the entire spectrum of initial conditions relevant to the perturbation's evolution. This procedure requires us to check two important conditions for each sampled value of $k/(a_0 H_0)$. First, we must verify that the Jeans instability is recovered for each solution of the system described by the equations \eqref{fullDelta} and \eqref{fullPhi}. Second, we need to ensure that the perturbative solution obtained from the linearized equations, as given by \eqref{PerturbativeAnsatzLHS} and \eqref{PerturbativeAnsatzRHS}, does not deviate significantly from the solution predicted by GR. To quantify the deviations from GR, we impose a heuristic criterion that the deviations in the matter power spectrum of quadratic gravity compared with GR should remain below $10\%$ at present. This threshold ensures that the quadratic theory does not introduce significant departures from GR in the late time limit, thereby maintaining consistency with current observational limits on deviations from standard gravitational theory. From the computation of the matter power spectrum, we observe that $P \propto \langle\tilde{\delta}(\boldsymbol{k})\tilde{\delta}^*(\boldsymbol{k})\rangle$ (where the tilde indicates the Fourier transform of the matter overdensity). Therefore, our constraints for the range of values that the ratio of the matter overdensities, between each theory, must range between $\sqrt{9/10}$ and $\sqrt{11/10}$. In FIG.~\ref{fig:ListPlotPointsSmallAlpha}, we plot the deviation between the scalar perturbations in quadratic gravity and general relativity, evaluated at present, against the choice of initial condition for $K_0=k/(a_0 H_0)$. Where the mode in quadratic gravity is approaching the GR limit $\bar{\alpha}=0$. The mode is approximated by the perturbative solution given by \eqref{PerturbativeAnsatzLHS}, such that $\delta^{\text{QG}}_m \equiv \delta^{\text{GR}}_m+\bar{\alpha} \delta^{(1)}_m + \mathcal{O}(\bar{\alpha}^2)$. We plot these results for $\bar{\alpha}=\{10^{-13},10^{-14}, 10^{-15} \}$. For $\bar{\alpha} =10^{-13}$, there is a high suppression of the perturbation in quadratic gravity compared to GR as $K_0$ increases, and consequently, the Jeans instability would be suppressed at these values. Based on the criterion that deviations from GR are $\leq 10\% $, we find that a weak upper bound of $\bar{\alpha}=\alpha H_0^2 \lesssim 10^{-14}$ is sufficient to satisfy both the requirement that structure growth is recovered and the constraint on deviations from GR for all sampled values of $K_0$.

\section{Conclusions} \label{Conclusions}

The framework of quadratic gravity provides an opportunity to resolve some of the limitations of GR, and can give rise to new phenomenology via the propagation of new degrees of freedom beyond GR and $\Lambda$CDM. Previous work has established that the condition $\alpha\geq 0$ is essential for preventing tachyonic instabilities in a Minkowski background. In this study, we extended the investigation to an expanding cosmological background during the era of matter domination, where we derive further constraints on the free parameter $\alpha$. Specifically, we find that $10^{14}H^2_0 \lesssim \alpha^{-1} \ll M^2_{\text{cutoff}}$ is necessary to maintain stability across various sectors of the theory. These constraints emerge from requiring non-negative $\alpha$ to eliminate classical instabilities in both the vector and tensor sectors, while also ensuring that deviations from GR remain sufficiently small and that the growth of perturbations within the scalar sector is recovered. These quadratic models of gravity deviate from GR at high-energy scales and may be suitable for studying early universe physics. Looking forward, future work may explore more stringent constraints on $\alpha$ in contexts where the UV completion of gravity plays a crucial role. This can be achieved through a variety of analytical approaches. One promising avenue involves examining the solutions of static and spherically symmetric, as well as rotating black holes, within these modified gravitational models. Such studies could uncover unique signatures of the modified dynamics in the strong-field regime, which may be absent in weak-field approximations. Additionally, investigating the behavior of degrees of freedom during inflation presents another fruitful path. This aspect has been previously discussed in the literature and could yield insights into how this class of models influences the early universe's dynamics. By probing these scenarios, one may better understand the implications of our findings and refine the bounds on 
$\alpha$.

The analysis in this work has been purely theoretical. Further bounds on the parameter space of these models may be found from observational constraints, particularly with data coming from gravitational wave catalogs.

\section{Acknowledgments}
BS acknowledges support from a Science and Technologies Facilities Council (STFC) Doctoral Training Grant. MS acknowledges support from the Science and Tech-
nology Facility Council (STFC), UK, under the research
grant ST/X000753/1.

\appendix
\section{Dynamical equations of motion for the scalar modes}

The scalar perturbation sector has two independent propagating degrees of freedom, the gauge-invariant fields $\delta_m$, related to the matter overdensities, and $\Phi$, one of the Bardeen potentials. They need to satisfy the following equations of motion:
\begin{align}
\delta_{\mathrm{m}}''&=\left(\frac{\left(3 w +3\right) \Omega}{2}+6 w -3 c_{s}^{2}-2\right) \delta_{\mathrm{m}}' +\frac{9 K^{2} \left(1+w \right) Y \Phi'}{2 K^{4} \bar\alpha  +9 Y}\nonumber\\
&+\frac{\delta_{\mathrm{m}}}{4 K^{4} \bar\alpha  +18 Y}\left[9 \left(\Omega  \left(1+w \right)-2 c_{s}^{2}\right) Y \,K^{2}-4 K^{6} c_{s}^{2} \bar\alpha  -18 \bar\alpha  \left(\left(1+w \right) \left(w-c_{s}^{2} \right) \Omega +2 c_{s}^{4}-\frac{10 w}{3}+\frac{4 c_{s}^{2}}{3}-2 \eta \right) K^{4}\right.\nonumber\\
&-\left.81 Y \left(\left(1+w \right) \left(-c_{s}^{2}+w \right) \Omega +2 c_{s}^{4}-\frac{10 w}{3}+\frac{4 c_{s}^{2}}{3}-2 \eta \right)\right]-\frac{6 \Phi  \left(\bar\alpha  \left(c_{s}^{2}-\frac{1}{3}\right) K^{4}-\frac{K^{2} Y}{2}-\frac{9 [\Omega  (1+w)-2 c_{s}^{2}] Y}{4}\right) }{(2 K^{4} \bar\alpha  +9 Y) [K^{2} (1+w )]^{-1}}\label{fullDelta}\,,
\\
    \Phi''&=\frac{9 \Omega  Y \delta_{\mathrm{m}}'}{4 \bar\alpha  \,K^{4}}+\frac{\left(6 \left(K^{4} \bar\alpha  -\frac{9 Y}{2}\right) \left(1+w \right) \Omega -12 K^{4} \bar\alpha  +18 Y \right) \Phi'}{4 K^{4} \bar\alpha  +18 Y}\nonumber\\
    &+\frac{3 \Omega  \left(\bar\alpha^{2} \left(c_s^2 -\frac{1}{3}\right) K^{8}-\frac{K^{6} \bar\alpha  Y}{2}-\frac{9 \left(\Omega  \left(1+w \right)+w -2 c_s^2 -1\right) \bar\alpha  Y \,K^{4}}{2}-\frac{9 K^{2} Y^{2}}{4}-\frac{81 Y^{2} \left(w -c_s^2 \right)}{4}\right) \delta_{\mathrm{m}}}{2 \left(K^{4} \bar\alpha  +\frac{9 Y}{2}\right) K^{4} \bar\alpha }\nonumber\\
    &+\frac{\Phi}{8 \left(K^{4} \bar\alpha  +\frac{9 Y}{2}\right) K^{2} \bar\alpha }\left[36 \bar\alpha  \left(\bar\alpha  \left(c_s^2 -1\right) \left(1+w \right) \Omega +\frac{8 \bar\alpha }{9}-\frac{Y}{9}\right) K^{6}-8 \bar\alpha^{2} K^{8}-54 \bar\alpha  \left(\Omega  \left(1+w \right)-\frac{2}{9}\right) Y \,K^{4}\right.\nonumber\\
    &-\left.162 Y \left(\bar\alpha  \left(1+w \right)^{2} \Omega^{2}-\bar\alpha  \left(c_s^2 +1\right) \left(1+w \right) \Omega +\frac{Y}{9}\right) K^{2}-81 \Omega  \,Y^{2} \left(1+w \right)\right].\label{fullPhi}
\end{align}

The system described by \eqref{fullDelta}, \eqref{fullPhi}, can be rendered as a fourth-order ODE in terms of $\delta_m$.

\begin{align}
    \delta''''_m & + A\delta'''_m + \frac{B}{EF} \delta''_m + \frac{C}{EF} \delta'_m + \frac{2D}{E} \delta_m = 0  \label{FourthOrderSystem} \\ \text{Where} \;
    A=&-\frac{6 Y}{2 \bar{\alpha}  K^2+3 Y}+\frac{4 \bar{\alpha}  K^4-6 K^2 Y}{4 \bar{\alpha}  K^4+6 K^2 Y+27 Y \Omega }-9 \Omega +8 
\end{align}

\begin{align}
     \notag B&=  32 \bar{\alpha} ^3 K^8+8 \alpha ^2 K^6 (\bar{\alpha}  (27 (\Omega -4) \Omega +80)+14 Y)+24 \bar{\alpha} K^4 Y (\bar{\alpha}  (3 \Omega  (9 \Omega -23)+56)+5 Y)\\&+18 K^2 Y \left(3 \bar{\alpha} ^2
   \Omega  (3 \Omega  (9 \Omega -32)+68)+2 Y^2+\bar{\alpha}  (3 \Omega  (9 \Omega -8)+32) Y\right) \nonumber \\&+81 Y^2 \Omega  (\bar{\alpha}  (3 \Omega  (9 \Omega -20)+44)+2 Y)
\end{align}

\begin{align}
    \notag C &= 16\bar{\alpha} ^3 K^8 (4-3 \Omega )-8 \alpha ^2 K^6 (48 \bar{\alpha} (\Omega -1)+7 Y (3 \Omega -4))-12 \bar{\alpha}  K^4 Y (\bar{\alpha} (9 \Omega  (\Omega +6)-64)+5 Y (3 \Omega -4))\\&-18
   K^2 Y \left(-3 \bar{\alpha} ^2 \Omega  \left(9 \Omega ^2-48 \Omega +40\right)+(3 \Omega -4) Y^2+2 \bar{\alpha}  (3 \Omega -4) (3 \Omega +2) Y\right) \nonumber\\&+243 Y^2 \Omega  (\bar{\alpha}
   (3 (\Omega -4) \Omega +8)-Y (\Omega -2))
\end{align}

\begin{align}
        D&=K^2 \Omega  \left(4 \bar{\alpha} ^2 K^4-9 Y^2\right)
   \\
   E&=4 \bar{\alpha}  \left(4 \bar{\alpha}  K^4+6 K^2 Y+27 Y \Omega \right)
   \\
   F&=2 \bar{\alpha}  K^2+3 Y
\end{align}

\bibliography{mybib}

\begin{thebibliography}{15}%
\makeatletter
\providecommand \@ifxundefined [1]{%
 \@ifx{#1\undefined}
}%
\providecommand \@ifnum [1]{%
 \ifnum #1\expandafter \@firstoftwo
 \else \expandafter \@secondoftwo
 \fi
}%
\providecommand \@ifx [1]{%
 \ifx #1\expandafter \@firstoftwo
 \else \expandafter \@secondoftwo
 \fi
}%
\providecommand \natexlab [1]{#1}%
\providecommand \enquote  [1]{``#1''}%
\providecommand \bibnamefont  [1]{#1}%
\providecommand \bibfnamefont [1]{#1}%
\providecommand \citenamefont [1]{#1}%
\providecommand \href@noop [0]{\@secondoftwo}%
\providecommand \href [0]{\begingroup \@sanitize@url \@href}%
\providecommand \@href[1]{\@@startlink{#1}\@@href}%
\providecommand \@@href[1]{\endgroup#1\@@endlink}%
\providecommand \@sanitize@url [0]{\catcode `\\12\catcode `\$12\catcode `\&12\catcode `\#12\catcode `\^12\catcode `\_12\catcode `\%12\relax}%
\providecommand \@@startlink[1]{}%
\providecommand \@@endlink[0]{}%
\providecommand \url  [0]{\begingroup\@sanitize@url \@url }%
\providecommand \@url [1]{\endgroup\@href {#1}{\urlprefix }}%
\providecommand \urlprefix  [0]{URL }%
\providecommand \Eprint [0]{\href }%
\providecommand \doibase [0]{https://doi.org/}%
\providecommand \selectlanguage [0]{\@gobble}%
\providecommand \bibinfo  [0]{\@secondoftwo}%
\providecommand \bibfield  [0]{\@secondoftwo}%
\providecommand \translation [1]{[#1]}%
\providecommand \BibitemOpen [0]{}%
\providecommand \bibitemStop [0]{}%
\providecommand \bibitemNoStop [0]{.\EOS\space}%
\providecommand \EOS [0]{\spacefactor3000\relax}%
\providecommand \BibitemShut  [1]{\csname bibitem#1\endcsname}%
\let\auto@bib@innerbib\@empty
\bibitem [{\citenamefont {Clifton}\ \emph {et~al.}(2012)\citenamefont {Clifton}, \citenamefont {Ferreira}, \citenamefont {Padilla},\ and\ \citenamefont {Skordis}}]{Clifton:2011jh}%
  \BibitemOpen
  \bibfield  {author} {\bibinfo {author} {\bibfnamefont {T.}~\bibnamefont {Clifton}}, \bibinfo {author} {\bibfnamefont {P.~G.}\ \bibnamefont {Ferreira}}, \bibinfo {author} {\bibfnamefont {A.}~\bibnamefont {Padilla}},\ and\ \bibinfo {author} {\bibfnamefont {C.}~\bibnamefont {Skordis}},\ }\href {https://doi.org/10.1016/j.physrep.2012.01.001} {\bibfield  {journal} {\bibinfo  {journal} {Phys. Rept.}\ }\textbf {\bibinfo {volume} {513}},\ \bibinfo {pages} {1} (\bibinfo {year} {2012})},\ \Eprint {https://arxiv.org/abs/1106.2476} {arXiv:1106.2476 [astro-ph.CO]} \BibitemShut {NoStop}%
\bibitem [{\citenamefont {Stelle}(1978)}]{Stelle:1977ry}%
  \BibitemOpen
  \bibfield  {author} {\bibinfo {author} {\bibfnamefont {K.~S.}\ \bibnamefont {Stelle}},\ }\href {https://doi.org/10.1007/BF00760427} {\bibfield  {journal} {\bibinfo  {journal} {Gen. Rel. Grav.}\ }\textbf {\bibinfo {volume} {9}},\ \bibinfo {pages} {353} (\bibinfo {year} {1978})}\BibitemShut {NoStop}%
\bibitem [{\citenamefont {Fradkin}\ and\ \citenamefont {Tseytlin}(1982)}]{Fradkin:1981iu}%
  \BibitemOpen
  \bibfield  {author} {\bibinfo {author} {\bibfnamefont {E.~S.}\ \bibnamefont {Fradkin}}\ and\ \bibinfo {author} {\bibfnamefont {A.~A.}\ \bibnamefont {Tseytlin}},\ }\href {https://doi.org/10.1016/0550-3213(82)90444-8} {\bibfield  {journal} {\bibinfo  {journal} {Nucl. Phys. B}\ }\textbf {\bibinfo {volume} {201}},\ \bibinfo {pages} {469} (\bibinfo {year} {1982})}\BibitemShut {NoStop}%
\bibitem [{\citenamefont {Hindawi}\ \emph {et~al.}(1996)\citenamefont {Hindawi}, \citenamefont {Ovrut},\ and\ \citenamefont {Waldram}}]{Hindawi:1995an}%
  \BibitemOpen
  \bibfield  {author} {\bibinfo {author} {\bibfnamefont {A.}~\bibnamefont {Hindawi}}, \bibinfo {author} {\bibfnamefont {B.~A.}\ \bibnamefont {Ovrut}},\ and\ \bibinfo {author} {\bibfnamefont {D.}~\bibnamefont {Waldram}},\ }\href {https://doi.org/10.1103/PhysRevD.53.5583} {\bibfield  {journal} {\bibinfo  {journal} {Phys. Rev. D}\ }\textbf {\bibinfo {volume} {53}},\ \bibinfo {pages} {5583} (\bibinfo {year} {1996})},\ \Eprint {https://arxiv.org/abs/hep-th/9509142} {arXiv:hep-th/9509142} \BibitemShut {NoStop}%
\bibitem [{\citenamefont {Bogdanos}\ \emph {et~al.}(2010)\citenamefont {Bogdanos}, \citenamefont {Capozziello}, \citenamefont {De~Laurentis},\ and\ \citenamefont {Nesseris}}]{Bogdanos:2009tn}%
  \BibitemOpen
  \bibfield  {author} {\bibinfo {author} {\bibfnamefont {C.}~\bibnamefont {Bogdanos}}, \bibinfo {author} {\bibfnamefont {S.}~\bibnamefont {Capozziello}}, \bibinfo {author} {\bibfnamefont {M.}~\bibnamefont {De~Laurentis}},\ and\ \bibinfo {author} {\bibfnamefont {S.}~\bibnamefont {Nesseris}},\ }\href {https://doi.org/10.1016/j.astropartphys.2010.08.001} {\bibfield  {journal} {\bibinfo  {journal} {Astropart. Phys.}\ }\textbf {\bibinfo {volume} {34}},\ \bibinfo {pages} {236} (\bibinfo {year} {2010})},\ \Eprint {https://arxiv.org/abs/0911.3094} {arXiv:0911.3094 [gr-qc]} \BibitemShut {NoStop}%
\bibitem [{\citenamefont {Hinterbichler}\ and\ \citenamefont {Saravani}(2016)}]{Hinterbichler:2015soa}%
  \BibitemOpen
  \bibfield  {author} {\bibinfo {author} {\bibfnamefont {K.}~\bibnamefont {Hinterbichler}}\ and\ \bibinfo {author} {\bibfnamefont {M.}~\bibnamefont {Saravani}},\ }\href {https://doi.org/10.1103/PhysRevD.93.065006} {\bibfield  {journal} {\bibinfo  {journal} {Phys. Rev. D}\ }\textbf {\bibinfo {volume} {93}},\ \bibinfo {pages} {065006} (\bibinfo {year} {2016})},\ \Eprint {https://arxiv.org/abs/1508.02401} {arXiv:1508.02401 [hep-th]} \BibitemShut {NoStop}%
\bibitem [{\citenamefont {Nelson}\ \emph {et~al.}(2010)\citenamefont {Nelson}, \citenamefont {Ochoa},\ and\ \citenamefont {Sakellariadou}}]{Nelson:2010rt}%
  \BibitemOpen
  \bibfield  {author} {\bibinfo {author} {\bibfnamefont {W.}~\bibnamefont {Nelson}}, \bibinfo {author} {\bibfnamefont {J.}~\bibnamefont {Ochoa}},\ and\ \bibinfo {author} {\bibfnamefont {M.}~\bibnamefont {Sakellariadou}},\ }\href {https://doi.org/10.1103/PhysRevD.82.085021} {\bibfield  {journal} {\bibinfo  {journal} {Phys. Rev. D}\ }\textbf {\bibinfo {volume} {82}},\ \bibinfo {pages} {085021} (\bibinfo {year} {2010})},\ \Eprint {https://arxiv.org/abs/1005.4276} {arXiv:1005.4276 [hep-th]} \BibitemShut {NoStop}%
\bibitem [{\citenamefont {De~Felice}\ \emph {et~al.}(2023)\citenamefont {De~Felice}, \citenamefont {Kawaguchi}, \citenamefont {Mizui},\ and\ \citenamefont {Tsujikawa}}]{DeFelice:2023psw}%
  \BibitemOpen
  \bibfield  {author} {\bibinfo {author} {\bibfnamefont {A.}~\bibnamefont {De~Felice}}, \bibinfo {author} {\bibfnamefont {R.}~\bibnamefont {Kawaguchi}}, \bibinfo {author} {\bibfnamefont {K.}~\bibnamefont {Mizui}},\ and\ \bibinfo {author} {\bibfnamefont {S.}~\bibnamefont {Tsujikawa}},\ }\href {https://doi.org/10.1103/PhysRevD.108.123524} {\bibfield  {journal} {\bibinfo  {journal} {Phys. Rev. D}\ }\textbf {\bibinfo {volume} {108}},\ \bibinfo {pages} {123524} (\bibinfo {year} {2023})},\ \Eprint {https://arxiv.org/abs/2309.01835} {arXiv:2309.01835 [gr-qc]} \BibitemShut {NoStop}%
\bibitem [{\citenamefont {Anselmi}(2018)}]{Anselmi:2018kgz}%
  \BibitemOpen
  \bibfield  {author} {\bibinfo {author} {\bibfnamefont {D.}~\bibnamefont {Anselmi}},\ }\href {https://doi.org/10.1007/JHEP02(2018)141} {\bibfield  {journal} {\bibinfo  {journal} {JHEP}\ }\textbf {\bibinfo {volume} {02}},\ \bibinfo {pages} {141}},\ \Eprint {https://arxiv.org/abs/1801.00915} {arXiv:1801.00915 [hep-th]} \BibitemShut {NoStop}%
\bibitem [{\citenamefont {Mukhanov}\ \emph {et~al.}(1992)\citenamefont {Mukhanov}, \citenamefont {Feldman},\ and\ \citenamefont {Brandenberger}}]{Mukhanov:1990me}%
  \BibitemOpen
  \bibfield  {author} {\bibinfo {author} {\bibfnamefont {V.~F.}\ \bibnamefont {Mukhanov}}, \bibinfo {author} {\bibfnamefont {H.~A.}\ \bibnamefont {Feldman}},\ and\ \bibinfo {author} {\bibfnamefont {R.~H.}\ \bibnamefont {Brandenberger}},\ }\href {https://doi.org/10.1016/0370-1573(92)90044-Z} {\bibfield  {journal} {\bibinfo  {journal} {Phys. Rept.}\ }\textbf {\bibinfo {volume} {215}},\ \bibinfo {pages} {203} (\bibinfo {year} {1992})}\BibitemShut {NoStop}%
\bibitem [{\citenamefont {Bardeen}(1980)}]{Bardeen:1980kt}%
  \BibitemOpen
  \bibfield  {author} {\bibinfo {author} {\bibfnamefont {J.~M.}\ \bibnamefont {Bardeen}},\ }\href {https://doi.org/10.1103/PhysRevD.22.1882} {\bibfield  {journal} {\bibinfo  {journal} {Phys. Rev. D}\ }\textbf {\bibinfo {volume} {22}},\ \bibinfo {pages} {1882} (\bibinfo {year} {1980})}\BibitemShut {NoStop}%
\bibitem [{\citenamefont {Peebles}(1980)}]{Peebles:1980yev}%
  \BibitemOpen
  \bibfield  {author} {\bibinfo {author} {\bibfnamefont {P.~J.}\ \bibnamefont {Peebles}},\ }\href@noop {} {\emph {\bibinfo {title} {{The Large-Scale Structure of the Universe}}}}\ (\bibinfo  {publisher} {Princeton University Press},\ \bibinfo {year} {1980})\BibitemShut {NoStop}%
\bibitem [{\citenamefont {Lovelock}(1971)}]{Lovelock:1971yv}%
  \BibitemOpen
  \bibfield  {author} {\bibinfo {author} {\bibfnamefont {D.}~\bibnamefont {Lovelock}},\ }\href {https://doi.org/10.1063/1.1665613} {\bibfield  {journal} {\bibinfo  {journal} {J. Math. Phys.}\ }\textbf {\bibinfo {volume} {12}},\ \bibinfo {pages} {498} (\bibinfo {year} {1971})}\BibitemShut {NoStop}%
\bibitem [{\citenamefont {Arnowitt}\ \emph {et~al.}(2008)\citenamefont {Arnowitt}, \citenamefont {Deser},\ and\ \citenamefont {Misner}}]{Arnowitt:1962hi}%
  \BibitemOpen
  \bibfield  {author} {\bibinfo {author} {\bibfnamefont {R.~L.}\ \bibnamefont {Arnowitt}}, \bibinfo {author} {\bibfnamefont {S.}~\bibnamefont {Deser}},\ and\ \bibinfo {author} {\bibfnamefont {C.~W.}\ \bibnamefont {Misner}},\ }\href {https://doi.org/10.1007/s10714-008-0661-1} {\bibfield  {journal} {\bibinfo  {journal} {Gen. Rel. Grav.}\ }\textbf {\bibinfo {volume} {40}},\ \bibinfo {pages} {1997} (\bibinfo {year} {2008})},\ \Eprint {https://arxiv.org/abs/gr-qc/0405109} {arXiv:gr-qc/0405109} \BibitemShut {NoStop}%
\bibitem [{\citenamefont {Schutz}\ and\ \citenamefont {Sorkin}(1977)}]{Schutz:1977df}%
  \BibitemOpen
  \bibfield  {author} {\bibinfo {author} {\bibfnamefont {B.~F.}\ \bibnamefont {Schutz}}\ and\ \bibinfo {author} {\bibfnamefont {R.}~\bibnamefont {Sorkin}},\ }\href {https://doi.org/10.1016/0003-4916(77)90200-7} {\bibfield  {journal} {\bibinfo  {journal} {Annals Phys.}\ }\textbf {\bibinfo {volume} {107}},\ \bibinfo {pages} {1} (\bibinfo {year} {1977})}\BibitemShut {NoStop}%
\end{thebibliography}%


\end{document}